\documentstyle[aps]{revtex}
\input{epsf.tex}
\begin{document}
\draft
\title{Transition to the Fulde-Ferrel-Larkin-Ovchinnikov phases in three dimensions : 
a quasiclassical investigation at low temperature with Fourier expansion}
\author{C. Mora and R. Combescot}
\address{Laboratoire de Physique Statistique,
 Ecole Normale Sup\'erieure*,
24 rue Lhomond, 75231 Paris Cedex 05, France}
\date{Received \today}
\maketitle

\begin{abstract}
We investigate, in three spatial dimensions, the transition from the normal state to the Fulde-Ferrel-Larkin-Ovchinnikov superfluid phases.  We make use of a Fourier expansion for the order parameter and the Green's functions to handle the quasiclassical equations in the vicinity of the transition. We show that, below the tricritical point, the transition is always first order. We find that, at the transition, the higher Fourier components in the order parameter are always essentially negligible. Below the tricritical point we have the already known result that the order parameter has a spatial dependence which is essentially $\cos({\bf q}.{\bf r})$. However when the temperature is lowered, the order parameter switches to a sum of two cosines, with equal weigths and wavevector with the same length, but orthogonal directions. Finally by further lowering the temperature, and down to $T=0$, one finds a another transition toward an order parameter which is the sum of three cosines with again equal weigths and orthogonal directions. Hence the structure of the order parameter gets more complex as the temperature is lowered. On the other hand the resulting critical temperatures are found to be only slightly higher than the ones corresponding to the standard second order FFLO transition. We apply our results
to the specific case of ultracold Fermi gases and show that the differences in atomic populations of the two hyperfine states involved in the BCS condensation display sizeable variations when one goes from the normal state to the superfluid FFLO phases, or one FFLO phase to another. Experimentally this should allow to identify clearly the various phase transitions.
\end{abstract}

 
\section{INTRODUCTION}
Although initiated a long time ago \cite{ff,larkov} the problem of the Fulde-Ferrel-Larkin-Ovchinnikov phases is far from having found a complete solution. In its simplest version one disregards the standard coupling of the magnetic field to the orbital degrees of freedom and one looks at the coupling with the electronic spins which gives rise to a difference between the chemical potentials of the up and the down spin populations. The question is to know how the superconducting transition is modified under these circumstances, and naturally the answer is relevant to deal with the more complex structures expected to appear when the orbital coupling is taken into account and vortex-like structures will be produced \cite{klein}. A recent and quite interesting development has been the realization that this FFLO phases are equally of high interest for the physics of ultracold atomic gases in their superfluid BCS-like state \cite{bkk} , and also in the physics of quark matter and dense nuclear matter \cite{bowers,wilcz,canar} as it might be found in the center of neutron stars. Our results near the transition at higher temperature \cite{cm} have confirmed the earlier results found in the literature \cite{br,matsuo,buz1}. Namely the preferred order parameter is a one-dimensional order parameter, whether the space itself is three-dimensional (3D) or two-dimensional (2D) (as it is relevant for lamellar superconductors or cuprate superconductors). At the transition it is proportional to a simple cosine $\Delta ( {\bf  r}) \sim \cos({\bf  q}. {\bf  r})$, this result being exact in 2D and almost exact in 3D. However we have found recently \cite{mceuro} that, in 2D, the situation gets increasingly complex at low temperature. Indeed in this case the order parameter at the transition takes the form of a superposition of cosines, with wavevectors having the same length but different directions, the number of these cosines increasing indefinitely when the temperature is going to zero.

In the present paper our purpose is to explore the same problem in three dimensions. This question is technically much more difficult than in 2D, since the transition turns out to be always first order in contrast with the 2D case. This fact is already known from higher temperature results \cite{br,matsuo,buz1,cm} and we will find that it remains true down to $T=0$. This makes it impossible to make use of a Ginzburg-Landau type of expansion, which in 3D is valid only near the tricritical point and, in 2D, at any temperature. One is thus forced, even at the transition, to face the full non-linear problem, which is quite difficult to handle. A most convenient way to attack it is to make use of  the quasiclassical equations of Eilenberger \cite{eil1}, Larkin and 
Ovchinnikov \cite{larkov1}. Nevertheless these equations themselves are not so easy to manipulate, even numerically, with the prospect of finding possible solutions with a three-dimensional order parameter, in a way analogous to what we have done in 2D.
In a preceding paper \cite{qcplana} we have made a progress in this direction by showing that the introduction of a Fourier expansion 
in the quasiclassical equations allows to obtain a solution which 
converges very rapidly toward the exact result. As a consequence a few 
terms in the expansion provide an excellent approximation. In this preceding paper only the principle of this method and its application to the 
a transition with one-dimensional order parameter have been considered. It has been checked against results obtained by Matsuo et al \cite{matsuo}, which started also from the quasiclassical equations, but used another method to solve them. We have found excellent agreement. In the present paper we will apply our technique to more complex cases where the order parameter does not have anymore a simple one-dimensional dependence. Naturally this was actually the aim of introducing this Fourier expansion method. We will explore which order parameter is actually favored by FFLO phases, in three dimensions, at low temperature. We find that this is no longer a one-dimensional order parameter, as it found at higher temperature, with a spatial which is nearly given by $\Delta ( {\bf  r}) \sim \cos({\bf  q}. {\bf  r})$ at the transition. We will rather obtain an order parameter with a more complicated structure. At the transition it is nearly the sum of two or three cosines (this last case being found near $ T = 0$), with directions of oscillations orthogonal to each other. As a result the transition does no longer switch to a second order one at low temperature, as found by Matsuo et al \cite{matsuo}, and go to the second order transition originally investigated at $T = 0$ by Larkin and Ovchinnikov \cite{larkov}. It stays rather a first order transition down to $T= 0$. Our results have been briefly reported elsewhere \cite{cmeuro}. 

Finally we apply our results to the case of ultracold Fermi gases and calculate the difference between the atomic populations of the two hyperfine states involved in the formation of Cooper pairs. We find that these differences can be quite sizeable and allow experimentally a clear distinction between the various phases, normal or superfluid, which may be found in these systems. In particular this should allow experimentally a clear identification of the transitions between these phases.

\section{FORMALISM}

In this section we will briefly recall for completeness the main points of the method presented in Ref.\cite{qcplana}, hereafter referred to as I. However we will write it immediately for the most general case we have in mind, i.e. a general Fourier expansion for the order parameter $ \Delta ({\bf r})$ and the Green's functions. More explicit formulations can be found in I.

We write Eilenberger's equations for the diagonal $ g ( \omega , 
\hat{ {\bf  k} }, {\bf  r})$ and off-diagonal $ f ( \omega , \hat{ {\bf  k} 
}, {\bf  r})$ quasiclassical propagators. Here ${\bf r}$ is the spatial variable while $\hat{{\bf k}}$ is the angular location of the wavevector on the spherical Fermi surface of radius $ k_F$. Finally $\omega $ is a general Matsubara frequency, to be continued into $ \omega _n - i \bar{\mu }$, where $ \omega _n = (2 n + 1) \pi T$ is the standard Matsubara frequency and $2 \bar{\mu } = \mu 
_{\uparrow} - \mu_{\downarrow} $ is the chemical potential difference between spin up and spin down
populations of the particles forming Cooper pairs. These equations read  \cite{eil1}:
\begin{eqnarray}
( \omega + {\bf  k}.{\bf \nabla}) f ( \omega , \hat{ {\bf  k} }, {\bf  r}) 
= \Delta ( {\bf  r}) 
g ( \omega , \hat{ {\bf  k} }, {\bf  r}) \nonumber  \\
( \omega - {\bf  k}.{\bf \nabla}) f ^{+} ( \omega , \hat{ {\bf  k} }, {\bf  
r}) = \Delta ^{*}( {\bf  r}) 
g ( \omega , \hat{ {\bf  k} }, {\bf  r}) \nonumber  \\
2{\bf  k}.{\bf \nabla} g ( \omega , \hat{ {\bf  k} }, {\bf  r}) = \Delta 
^{*}( {\bf  r}) 
f ( \omega , \hat{ {\bf  k} }, {\bf  r}) -  \Delta ( {\bf  r}) 
f ^{+} ( \omega , \hat{ {\bf  k} }, {\bf  r})
\label{eqeilen}
\end{eqnarray}
in which we have simplified the writing by taking $ \hbar = 1$ and $ m = 1/2 $ for the particle mass. 
These linear equations are closed by the normalization 
condition :
\begin{eqnarray}
g ( \omega , \hat{ {\bf  k} }, {\bf  r}) = (1 - f ( \omega , \hat{ {\bf  k} 
}, {\bf  r}) 
f ^{+} ( \omega , \hat{ {\bf  k} }, {\bf  r}))^{1/2}
\label{eqnorm}
\end{eqnarray}
from which one can see that the last of Eq.(\ref{eqeilen}) results from the two first ones. Finally the order parameter itself is linked to $ f ( \omega , \hat{ {\bf  k} }, {\bf  r})$ by the self-consistency equation 
\cite{eil1,qcplana} which we will not write explicitely here.

We assume then a very general Fourier expansion form for the order parameter:
\begin{eqnarray}
\Delta ({\bf r}) =  \sum_{\{n_i\}}  \Delta_{\{n_i\}} \exp( i \sum_{i}n_{i}{\bf q}_{i}.{\bf r})
\label{dgen}
\end{eqnarray}
where the $n_i$ are relative integers with $i=1,2,...,N_q$ and the ${\bf q}_i$ are $N_q$ general wavevectors for which no specific relations are assumed. If we had only three wavevectors ${\bf q}_i$ with $i=1,2,3$ this would mean that we consider a general periodic order parameter. However we do not restrict ourselves to three wavevectors, and we will indeed consider below explicitely the case of four wavevectors. We note that such an order parameter is not periodic in general. Naturally the numerics gets rapidly more difficult when the number of wavevectors ${\bf q}_i$ increases. We will nevertheless assume for simplicity a real order parameter since, at least in three spatial dimensions, all the order parameters found for actual FFLO phases have this property. This implies $\Delta^{*} _{\{n_i\}} = \Delta_{\{-n_i\}}$. We also restrict ourselves as before to an order parameter having real components $\Delta_{\{n_i\}}$. Finally we will only explore the case where all the wavevectors  have equal length $|{\bf q}_i | = q$. Indeed we know from the original work \cite{larkov} of Larkin and Ovchinnikov that the physics of the FFLO phases is that there is a competition between the different directions of the wavevectors having same modulus. At least this is the prevalent situation near the transition, in which we are interested in.

We make for the Green's functions similar expansions:
\begin{eqnarray}
f({\bf r}) = \sum_{\{n_i\}} f _{\{n_i\}} \exp( i\sum_{i}n_{i}{\bf q}_{i}.{\bf r})	\hspace{7mm} &
f^{+}({\bf r}) = \sum_{\{n_i\}} f^{+} _{\{n_i\}} \exp( i\sum_{i}n_{i}{\bf q}_{i}.{\bf r})
& \hspace{7mm}  
g({\bf r}) = \sum_{\{n_i\}} g _{\{n_i\}}  \exp( i\sum_{i}n_{i}{\bf q}_{i}.{\bf r})
\label{eq11}
\end{eqnarray}
and substitute them into Eilenberger's equations Eq.(\ref{eqeilen}). We introduce for convenience
$ d _{\{n_i\}} = (f _{\{n_i\}} -  f^{+}_{\{n_i\}})/2i$, which gives $ f _{\{n_i\}} = (i - 
\omega /\kappa (\{n_i\}) ) d _{\{n_i\}}$, where $\kappa (\{n_i\})= {\bf k}. \sum_{i}n_{i}{\bf q}_{i}$. This leads to:
\begin{eqnarray}
d _{\{n_i\}} = - \frac{\kappa (\{n_i\})}{ \omega ^{2}+ \kappa ^{2} (\{n_i\})}
 \sum_{p_i=1}^{\infty} \Delta_{ \{p_i\}} (g _{ \{n_i - p_i\}}+ g _{\{n_i + p_i\}})   \nonumber  \\
g _{\{n_i\}} = \frac{1 }{\kappa (\{n_i\})} \sum_{p_i=1}^{\infty} \Delta_{ \{p_i\}} (d _{\{n_i - p_i\}}+ d _{\{n_i + p_i\}})
\label{recurgen}
\end{eqnarray}
Just as in I, we consider only order parameters with zero spatial average i.e. $\Delta_{\{0\}}=0$ when all the $n_i=0$, because mixing in the uniform BCS phase order parameter is likely to be unfavorable. Then it can be seen iteratively, as in I, that only odd components of $ d _{\{n_i\}}$ and even components of $ g _{\{n_i\}}$ are non zero, i.e. precisely $ d _{\{n_i\}} = 0$ if $  \sum_{i} n_i$ is even, and $ g _{\{n_i\}} = 0$ if $  \sum_{i} n_i$ is odd.

Eq.(\ref{recurgen}) look quite complicated, but it turned out that we had to deal with them only in rather simple cases. Indeed we studied first the case of two cosines, and concluded as in I that the weight of higher order harmonics for the order parameter is quite small (we will give specific results below). We have then taken this conclusion for granted when we studied three or four cosines. This means that we have mostly dealed with Eq.(\ref{recurgen}) only in the case where all the $\Delta_{\{n_i\}}$ are zero, except when $ \sum_{i} | n_i |=1$, i.e. all the $n_i$ are zero except one of them which is $\pm 1$. In this case we have only $2 N_q$ terms in the sums in the right-hand side of Eq.(\ref{recurgen}). Similarly we have studied in some cases for two cosines the situation where the weights $\Delta_{\{n_i\}}$ are not equal and we have come with the conclusion that the symmetric situation with equal weights is the most favorable. We have then taken this conclusion for granted in the rest of our study. This means finally that, in all these cases, we have to deal with a single non zero component of $\Delta_{\{n_i\}}$, which we call $ \Delta _1$ in the following. Naturally, apart from the complexity of the equations, the essential difficulty in this FFLO problem is that the functional space to be explored is huge, to say the least, and that it seems hopeless to explore it completely. One has thus to appeal to symmetry arguments, like the equal weight argument, to restrict it.

Our two sets of Fourier components, $ g _{\{n_i\}}$ and $ d _{\{n_i\}}$, are defined on a discrete $N_q$-dimensional lattice. However since on any site one of these two components is zero depending on the parity of $  \sum_{i} n_i$, as we have seen, we have actually to deal on each site with a single component. Then Eq.(\ref{recurgen}) relate the value of this component on a given site to its a value on the nearest neighbours of this site on the $N_q$-dimensional lattice. By the same kind of arguments as in I it can be seen that the Fourier components decrease very rapidly on the site goes far away from the origin. Hence we will find an approximate solution for our problem by requiring that our components are zero when the site is far away from the origin. Specifically we will take them to be zero when $ \sum_{i} | n_i | > N_{max}$. This serves as boundary condition for our linear system Eq.(\ref{recurgen}). In this way equations become a finite dimensional linear system, which we solve numerically by standard and efficient methods. In principle when we let $N_{max}$ go to $\infty$ we find the exact solution. In practice, as will be discussed below, the convergence is found to fast enough so that the size of our linear system stays quite reasonable. We note that, because we have the properties $ g _{\{-n_i\}} =  g _{\{n_i\}}$ and $ d _{\{-n_i\}} = - d _{\{n_i\}}$, we have only to deal with half of the sites satisfying $ \sum_{i} | n_i | \le N_{max}$. This domain $\mathcal{D}$ for summation over the ${\{n_i\}}$ is shown explicitely in Fig. 1 in the case of two cosines $N_q = 2$ and for $N_{max} = 3$ .
\begin{figure}[htbp]
\begin{center}
\vbox to 60mm{\hspace{0mm} \epsfysize=60mm \epsfbox{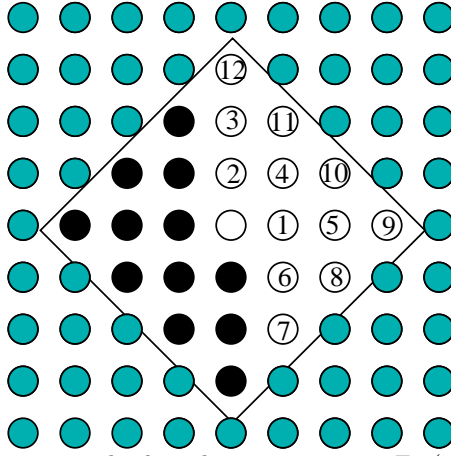}}
\caption{Lattice sites, in the $\{n_1,n_2\}$ space, involved in the summation in Eq.(\ref{recurgen1}),  in the case of two cosines $N_q = 2$ and for $N_{max} = 3$ . The numbers correspond to the actual ordering we use in writing the corresponding equations, which goes by increasing $| n_1 | + | n_2| $.}
\label{default}
\end{center}
\end{figure}
In practice we have set, as in I:
\begin{eqnarray}
d_{\{n_i\}} = \kappa(\{n_i\}) D_{\{n_i\}} g_0 \nonumber \\ 
g_{\{n_i\}} = G_{\{n_i\}} g_0 
\label{eqa}
\end{eqnarray}
where $g_0$ is for $g_{\{n_i=0\}}$, i.e. the value of $g_{\{n_i\}}$ at the origin of our discrete lattice. Then Eq.(\ref{recurgen}) becomes explicitely:
\begin{eqnarray}
[ \omega^2 + \kappa^2(\{n_i\}) ]
D_{\{n_i\}} + \Delta_1 \sum_{\{n_j\}} G_{\{n_i\}}  = 0 \nonumber \\
- \kappa(\{n_i\}) G_{\{n_i\}} + \Delta_1 \sum_{\{n_j\}} 
\kappa(\{n_j\})  D_{\{n_j\}} = 0
\label{recurgen1}
\end{eqnarray}
We normalize these new components by the condition $G_0 = 1$. When the normalization condition Eq.(\ref{eqnorm}) is expressed with these components, one finds:
\begin{eqnarray}
g^{-2}_0 = 1 + 2 \sum_{\mathcal{D}} G_{\{n_i\}}^2 + 2  \sum_{\mathcal{D}} 
[\omega^2-\kappa^2(\{n_i\})] D_{\{n_i\}}^2
\label{eqg0}
\end{eqnarray}
This component $g_0$, the spatial average of the Green's function, is just the quantity we need to
calculate \cite{qcplana} the free energy difference $ \Omega _{s} - \Omega _{n}$ between the superfluid and the normal phase:
\begin{eqnarray}
\frac{ \Omega _{s} - \Omega _{n}}{2 N _{0}} =  N_{q} \Delta_1^2 \ln [\frac{T}{T 
_{sp}(\bar{\mu }/T)}]  + 2 \pi  T  \sum_{n=0}^{ \infty} \int_{\omega _{n}}^{ \infty}
\, d \omega  \:  \int_{S_F} \! \! d\Omega_{{\bf k}} {\mathrm{Re}} \, [ g _{0} ( \omega - i \bar{\mu } , 
\hat{ {\bf  k} })
- 1 + \frac{N_{q} \Delta_1^2}{2 (\omega - i \bar{\mu }) ^{2}} ] 
\label{frenerg}
\end{eqnarray}
where $N _{0}$ is the single spin density of states at the Fermi surface. In Eq.(\ref{frenerg}) the angular average is over the direction of ${\bf {\hat k}}$ at the Fermi surface $S_F$. We have also, instead of the standard BCS critical temperature $T_{c0}$ at zero effective field $ {\bar \mu }=0$, introduced the critical temperature $T_{sp}(\bar{\mu }/T)$ for the spinodal transition, which is the second order transition from the normal to the uniform BCS phase. This $T_{sp}$ is conveniently seen as a function of $\bar{\mu }/T$, which is a kind of angular coordinate when one explores the ($\bar{\mu },T$) plane. More specifically, since we are only interested in the location $T_c$ of the first order transition, we look at fixed $\bar{\mu }/T$ for the highest temperature $T=T_c$ at which we will have $ \Omega _{s} - \Omega _{n}=0$. Naturally we maximize this temperature with respect to the amplitude $\Delta _1$ of the order parameter and to the common length $q$ of the wavevector ${\bf q}_i$.

\section{NUMERICAL RESULTS}

We start the presentation of our numerical results by the case of two wavevectors $N_q = 2$. Indeed as already mentionned we have in this case made specific explorations, the results of which we have taken for granted in the numerically more heavy situations of three or four wavevectors.

\subsection{Two wavevectors}

\begin{figure}[htbp]
\begin{center}
\vbox to 60mm{\hspace{0mm} \epsfysize=60mm \epsfbox{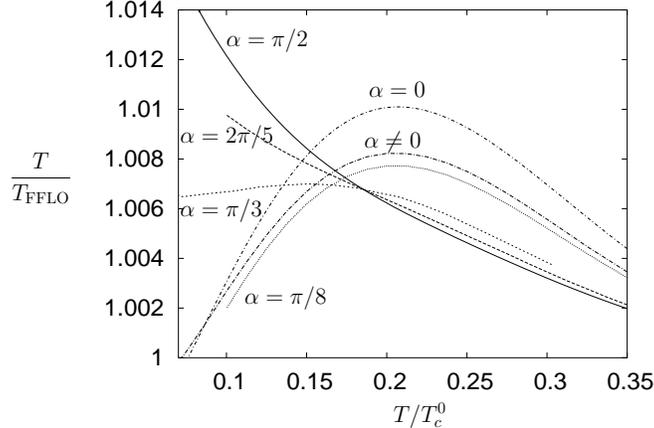}}
\caption{Critical temperature $T$ for the order parameter $\Delta ({\bf r}) = 2 \Delta_1 [ \cos({\bf q}_1 \cdot {\bf r})  + \cos({\bf q}_2 \cdot {\bf r}) ]$, for a selection of angles $\alpha $ between the wavevectors ${\bf q}_1$ and ${\bf q}_2$. Precisely we plot the ratio of the critical temperature $T/T_{ FFLO}$ to the FFLO critical temperature $T_{ FFLO}( \bar{\mu } / T)$ obtained for the same value of the ratio $  \bar{\mu } / T$. On the x-axis, instead of $  \bar{\mu } / T$, we give the value of $T_{ FFLO}( \bar{\mu } / T)$ itself, compared to the standard BCS critical temperature $T_{ c0} $.}
\label{figangl}
\end{center}
\end{figure}

We have investigated how the critical temperature $T$ depends on the angle between the two wavevectors ${\bf q}_1$ and ${\bf q}_2$. In this study we have assumed that only the lowest Fourier components are important, so that our order parameter is actually the sum of two cosines corresponding to these wavevectors. Our results are given in Fig.\ref{figangl}. As in I, we plot the ratio of the critical temperature $T/T_{ FFLO}$ to the FFLO critical temperature $T_{ FFLO}( \bar{\mu } / T)$ obtained for the same value of the ratio $  \bar{\mu } / T$. Instead of $  \bar{\mu } / T$, we give on the x-axis the value of $T_{ FFLO}( \bar{\mu } / T)$ itself, compared to the standard BCS critical temperature $T_{ c0} $.
We give results only for selected angles between $0$ and $\pi /2$, since the results are the same for $\alpha $ and $\pi -\alpha $. We also give the result for $\alpha =0$ which is just the result for a single cosine. It can be seen that, above $T_{ FFLO}/T_{ c0} \simeq 0.154 $, this is this single cosine order parameter which is favored. Below this temperature the order parameter switches discontinuously to two cosines, with an angle $\alpha = \pi /2$ between the corresponding wavevectors. Hence there is no situation where an angle $ 0 < \alpha  < \pi /2$ is favored. We note that the limit $ \alpha \rightarrow 0$ is singular. This can be seen on Fig.\ref{figangl} where this limit is noted $ \alpha \neq 0$. This singularity can be understood since, for two cosines, the spatial average of the square of the order parameter is always $ 4 \Delta _{1}^{2}$, even if the angle between the two vectors is very small, while it is $ 8 \Delta _{1}^{2}$ when this angle is exactly zero. In the following we take this symmetry result for granted and assume that the situation with orthogonal wavevectors is always the most favorable. We have also made some exploration of the evolution of the critical temperature when the relative weights $ \Delta 
_{i}$ of the cosines vary. For example, we have kept the 
wavevectors orthogonal or with an angle $ \pi /3 $ and minimized 
independently the two weights. We have found that, in either case, the 
optimum is for equal weights. 

\begin{figure}[htbp]
\begin{center}
\vbox to 60mm{\hspace{0mm} \epsfysize=60mm \epsfbox{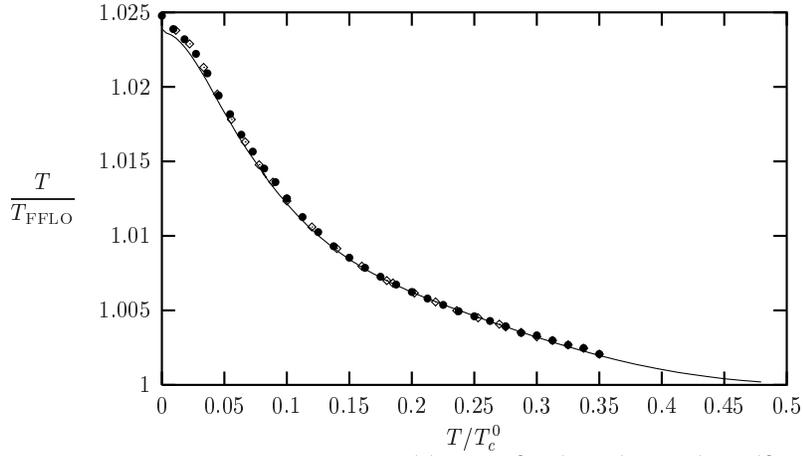}}
\caption{Critical temperature $T$ for the order parameter $\Delta ({\bf r}) = 2 \Delta_1 [ \cos({\bf q}_1 \cdot {\bf r})  + \cos({\bf q}_2 \cdot {\bf r}) ]$, for an angle $\alpha =\pi /2$ between the wavevectors ${\bf q}_1$ and ${\bf q}_2$. The full line is for $N_{max}=3$, while the filled circles and the open diamonds are for $N_{max}=4$ and $5$. Actually these last results $N_{max}=4$ and $5$ can not be distinguished within our precision.}
\label{fignmax}
\end{center}
\end{figure}

\begin{figure}[htbp]
\begin{center}
\vbox to 60mm{\hspace{0mm} \epsfysize=60mm \epsfbox{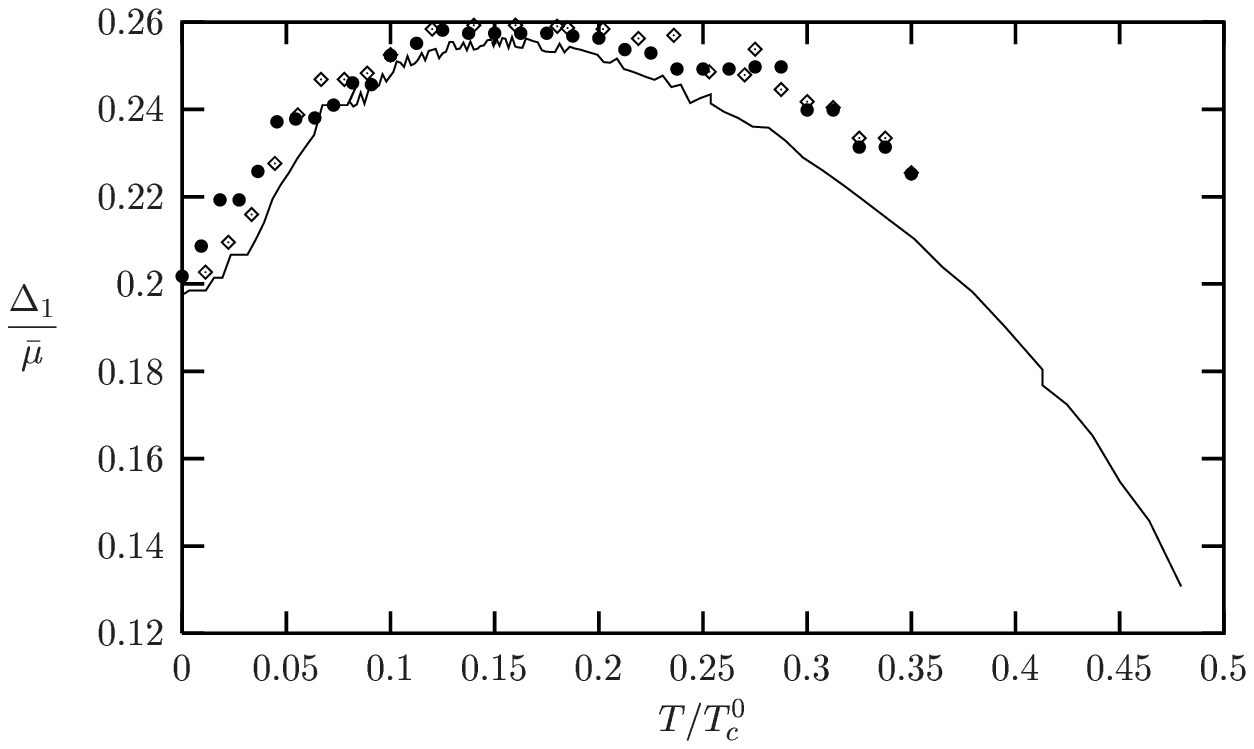}}
\vbox to 60mm{\hspace{0mm} \epsfysize=60mm \epsfbox{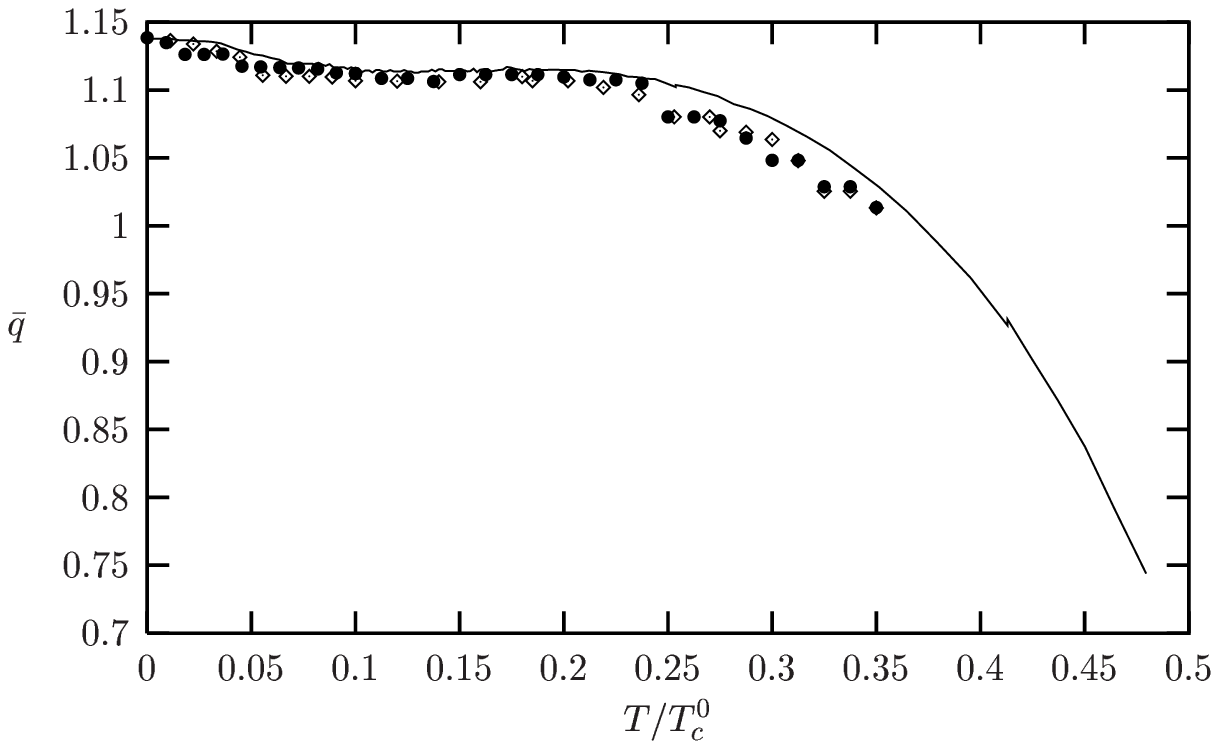}}
\caption{Relative amplitude $\Delta _{1}/{\bar \mu }$ of the order parameter and corresponding length $ {\bar q}=  q k _{F}/(2m \bar{\mu })$ of the wavevectors, as a function of the upper bound $N_{max}$ for the order of Fourier components. The full line is for $N_{max}=3$, the filled circles for $N_{max}=4$ and the open diamonds for $N_{max}=5$.}
\label{fignmaxdq}
\end{center}
\end{figure}

\begin{figure}[htbp]
\begin{center}
\vbox to 60mm{\hspace{0mm} \epsfysize=60mm \epsfbox{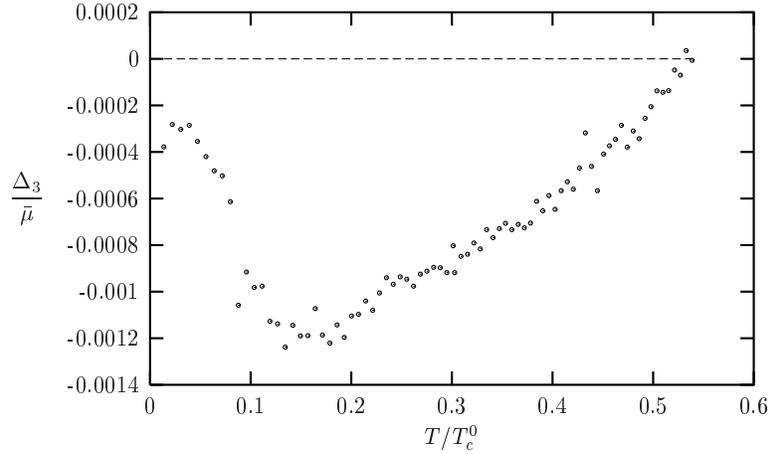}}
\caption{Relative amplitude of the higher harmonic $\Delta _{3}/{\bar \mu }$ for the order parameter Eq.\ref{eq2cosharm}.}
\label{figharm}
\end{center}
\end{figure}

We consider next the sensitivity to our upper bound $N_{max}$ to the order of Fourier components we take into account, that is we study how fast our procedure converges to the exact result, obtained in principle for $N_{max}=\infty$. We make this study only for the most favorable situation $\alpha = \pi /2$, which is the only one we consider from now on. The result is displayed in Fig.\ref{fignmax}, where we plot, as a function of the temperature, the results for $N_{max}=3$, $4$ and $5$. The results for $N_{max}=4$ and $N_{max}=5$ are undistinguishable within our precision, and it can be seen that $N_{max}=3$ gives already an excellent precision. This is actually better than what we have found in I, for the case of a single cosine. On the other hand the numerics is more complicated in the case of two cosines because the angular average requires two angular integrations, one on the polar angle and the other one on the azimuthal angle, while for symmetry reasons only the polar integration is necessary for a single cosine. We note that, in this respect, the situation does not get worse when we consider three or four cosines, since we have still only two angular integrations to perform.
We also give for completeness in Fig.\ref{fignmaxdq} the corresponding results for the size $\Delta _{1}/{\bar \mu }$ of the order parameter and the corresponding length $ {\bar q}=  q k _{F}/(2m \bar{\mu })$ of the wavevectors. The results are somewhat scattered, since our minimization procedure makes it difficult to find exactly the location of the minimum. On the other hand the critical temperature is much more precisely known.
Finally we consider how our results are modified when we go beyond the lowest harmonic approximation. We take for this purpose the order parameter:
\begin{eqnarray}
\Delta({\bf r}) = 2 \Delta_1 [ \cos({\bf q}_1 \cdot {\bf r})  + 
\cos({\bf q}_2 \cdot {\bf r})] +
2 \Delta_3 [ \cos(3{\bf q}_1 \cdot {\bf r})  + 
\cos(3 {\bf q}_2 \cdot {\bf r})]
\label{eq2cosharm}
\end{eqnarray}
and study the size of the higher harmonic correction $\Delta _{3}$ as a function of temperature. This is done for Nmax=4. The result is found in Fig.\ref{figharm}. We see that the harmonic $\Delta _{3}$ is always quite small, since the ratio $\Delta _{3}/\Delta _{1}$ is typically a few $10^{-3}$. Hence, just as in the case of the one-dimensional order parameter \cite{qcplana} , it is quite justified to consider only the lowest harmonic.
 We give also in Fig.\ref{figharm1} the results for the critical temperature, the lowest harmonic amplitude and the wavevector length with and without taking into account the higher harmonic $\Delta_{3}$. One can see that the results are indeed essentially undistinguishable.
\begin{figure}[htbp]
\begin{center}
\vbox to 60mm{\hspace{0mm} \epsfysize=60mm \epsfbox{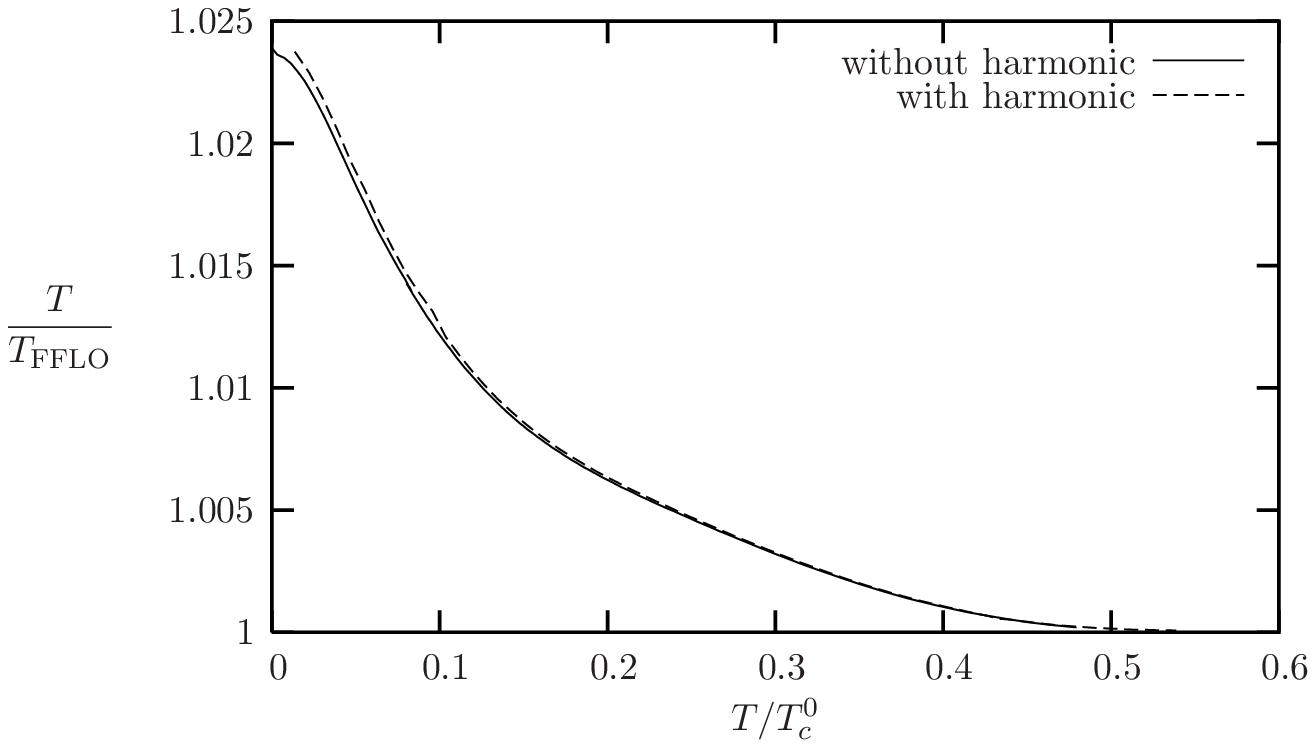}}
\label{}
\end{center}
\end{figure}
\begin{figure}[htbp]
\begin{center}
\vbox to 60mm{\hspace{0mm} \epsfysize=60mm \epsfbox{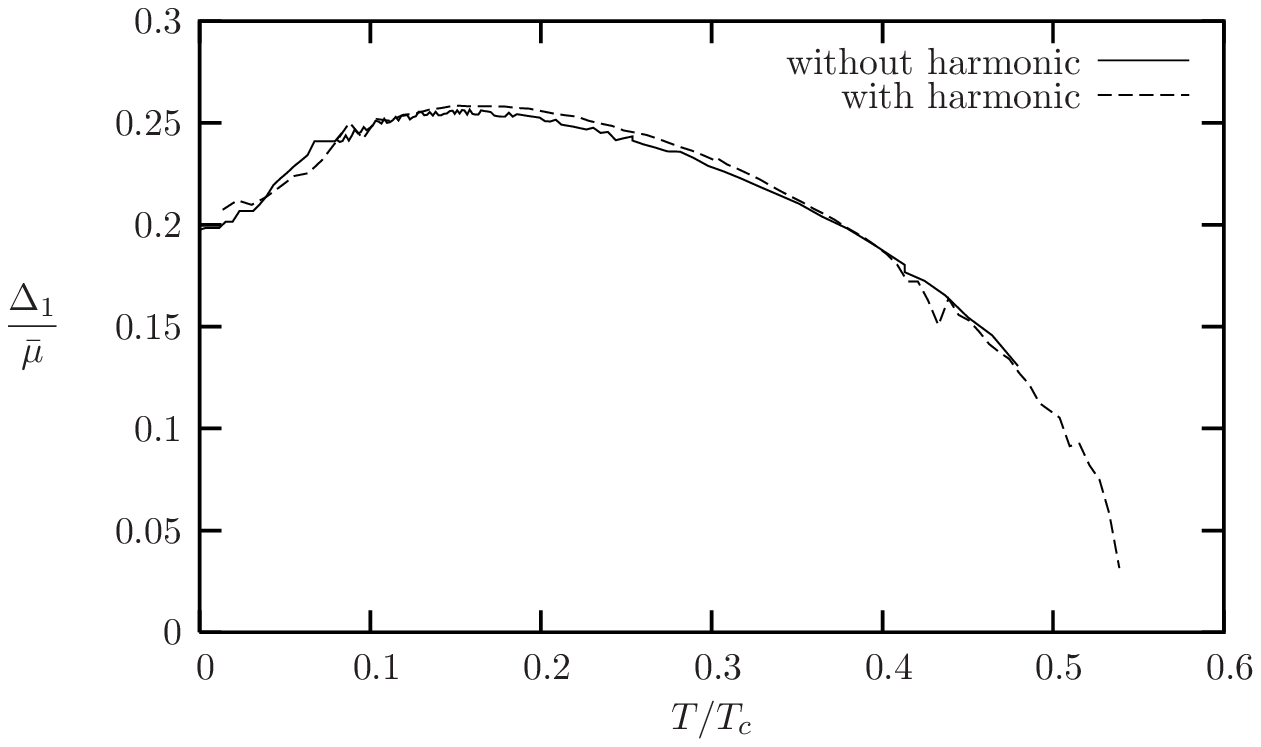}}
\vbox to 60mm{\hspace{0mm} \epsfysize=60mm \epsfbox{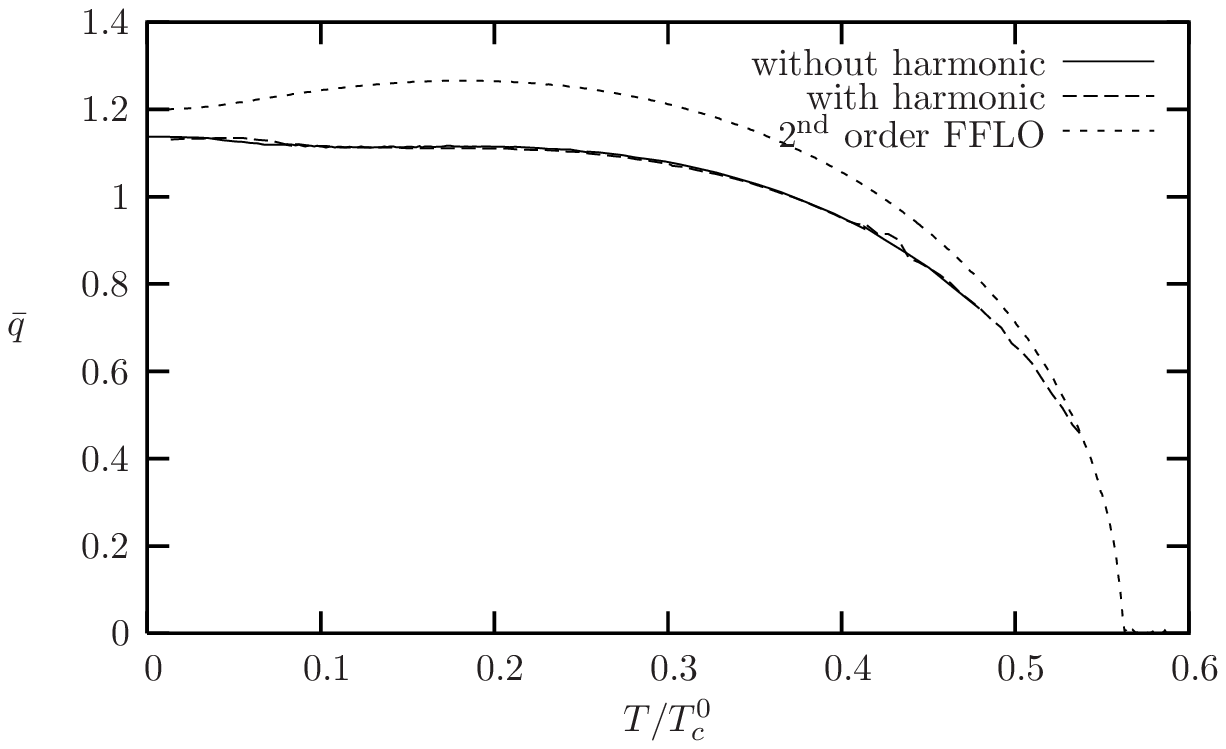}}
\caption{Critical temperature, lowest harmonic amplitude $\Delta _{1}/{\bar \mu }$ and corresponding wavevector length $ {\bar q}=  q k _{F}/(2m \bar{\mu })$ for the order parameter Eq.\ref{eq2cosharm} with $\Delta_{3}=0$ and $\Delta_{3} \neq 0$ (as given by Fig.\ref{figharm}) respectively.}
\label{figharm1}
\end{center}
\end{figure}

\subsection{Three wavevectors}
We consider now the case of three wavevectors ${\bf q}_1$, ${\bf q}_2$ and ${\bf q}_3$. Following the above study, we take for granted that higher harmonics give a negligible contribution so that we can restrict ourselves to the order parameter:
\begin{eqnarray}
\Delta({\bf r}) = 2 \Delta_1 [ \cos({\bf q}_1 \cdot {\bf r})  + 
\cos({\bf q}_2 \cdot {\bf r}) + \cos({\bf q}_3 \cdot {\bf r}) ]
\label{eq3cos}
\end{eqnarray}
which implies that we also assume the amplitude for the three cosines are equal. Similarly we assume
that the optimum order parameter corresponds to wavevectors with same length $|{\bf q}_1|=|{\bf q}_2|=|{\bf q}_3|$. Finally we have again studied in this case how the critical temperature depends on the angle between the three wavevectors, by taking them in a rhomboedral geometry and varying the rhomboedric angle. We have found again that the most favorable situation is found when the wavevectors are orthogonal. In the same way as for two wavevectors, this result is consistent with a phenomenological interpretation in terms of an effective repulsive potential between any two wavevectors, decreasing when the angle between them increases. With such a description the total potential is clearly minimized when the three wavevectors are orthogonal.

\begin{figure}[htbp]
\begin{center}
\vbox to 60mm{\hspace{0mm} \epsfysize=60mm \epsfbox{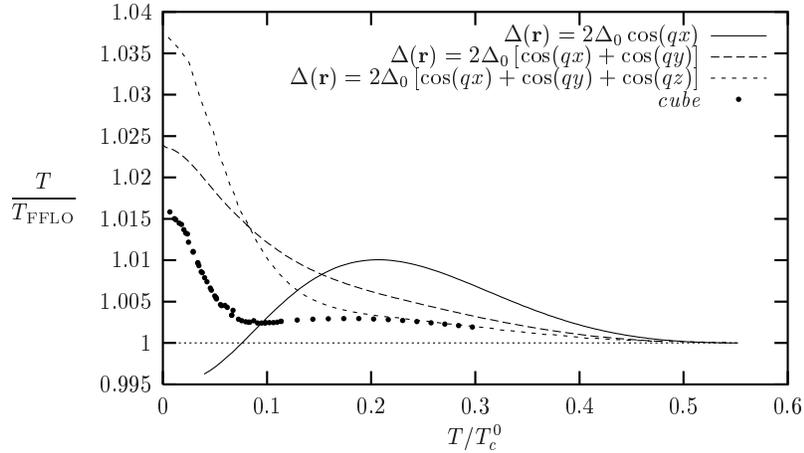}}
\caption{Critical temperature for an order parameter with respectively three cosines (short dashed), two cosines (long dashed) and a single cosine. The actual transition between the normal and the superfluid state corresponds to the highest possible transition temperature between these three possibilities. The filled dots refer to the "cube" configuration, considered in the next subsection.}
\label{figtcfincub}
\end{center}
\end{figure}

\begin{figure}[htbp]
\begin{center}
\vbox to 60mm{\hspace{0mm} \epsfysize=60mm \epsfbox{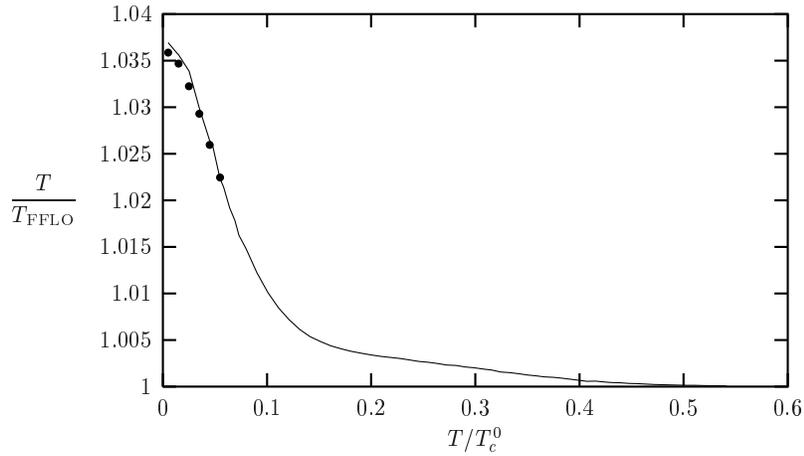}}
\caption{Critical temperature for an order parameter with three cosines. The full line is for $N_{max}=3$, while the filled circles are for $N_{max}=4$.}
\label{fig3cosnmax}
\end{center}
\end{figure}

\begin{figure}[htbp]
\begin{center}
\vbox to 60mm{\hspace{0mm} \epsfysize=60mm \epsfbox{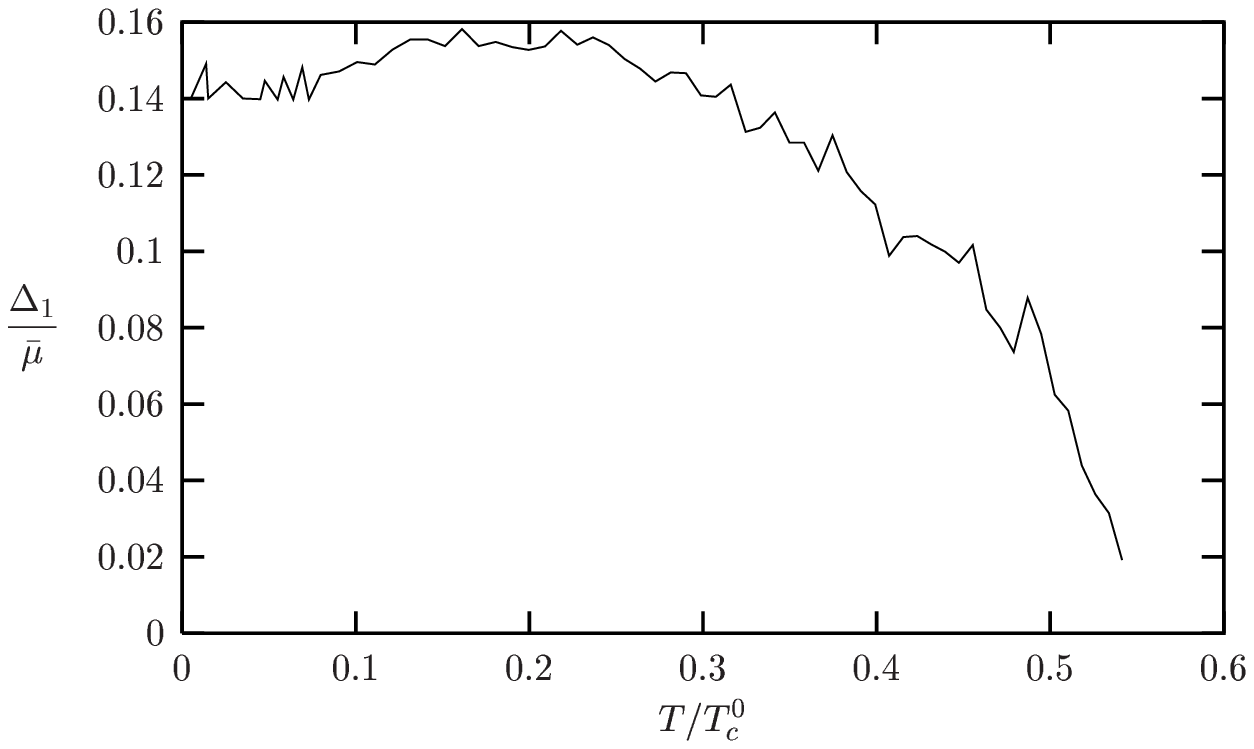}}
\vbox to 60mm{\hspace{0mm} \epsfysize=60mm \epsfbox{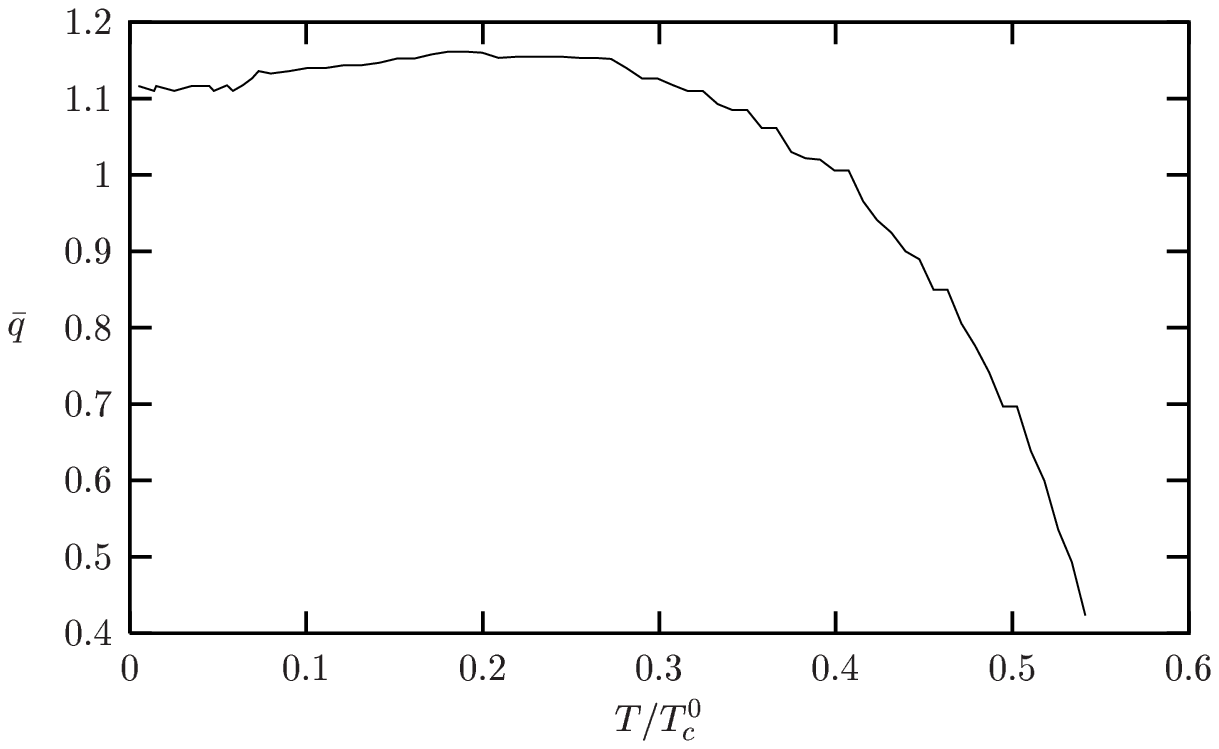}}
\caption{Order parameter amplitude $\Delta _{1}/{\bar \mu }$ and corresponding reduced wavevector length ${\bar q}$ as a function of temperature.}
\label{fig3cosdq}
\end{center}
\end{figure}

We present now our numerical results for this specific situation of orthogonal wavevectors. The critical temperature is given in Fig.\ref{figtcfincub}. We display again for comparison on this figure the results for a single cosine and for two cosines, given in Fig.\ref{figangl}. We see from this figure that the order parameter at the transition switches from a single cosine for $T_{ FFLO}/T_{ c0} > 0.154$ to an order parameter with two cosines for $0.080 < T_{ FFLO}/T_{ c0} < 0.154$, and to an order parameter with three cosines for $T_{ FFLO}/T_{ c0} < 0.080$. Naturally,  since they do not have the same symmetry, this implies that, inside the superfluid phase, there are first order transition lines between these various order parameters (or rather their continuation, because higher harmonics are expected to be more important deeper in the superfluid phase). On the other hand it is rather striking that the switch of the order parameter from a given symmetry to another has very little effect on the value of the critical temperature itself since $T_c/T_{ FFLO}$ is only slightly larger than unity.
The results in Fig.\ref{figtcfincub} for three cosines have been obtained for $N_{max}=3$. We have made several checks, by performing calculations for $N_{max}=4$, that convergence is indeed already reached. This is shown in Fig.\ref{fig3cosnmax} where the results for $N_{max}=4$ are barely distinguishable from those for $N_{max}=3$.
Finally we present for completeness in Fig.\ref{fig3cosdq} our corresponding results for the relative amplitude $\Delta _{1}/{\bar \mu }$ and the reduced wavevector length ${\bar q}$ of the order parameter.

\subsection{Four wavevectors}
As we have seen above, when $ T \rightarrow 0$, increasing the number of cosines from a single cosine to three increases the critical temperature. It is natural to wonder if this trend does not keep going for a larger number of cosines, all the more since this is indeed what happens in two dimensions \cite{mceuro} when $ T \rightarrow 0$. In order to explore this possibility we have considered the transition toward an order parameter which is the sum of four cosines:
\begin{eqnarray}
\Delta({\bf r}) = 2 \Delta_1 [ \cos({\bf q}_1 \cdot {\bf r})  + 
\cos({\bf q}_2 \cdot {\bf r}) + \cos({\bf q}_3 \cdot {\bf r}) + \cos({\bf q}_4 \cdot {\bf r})  ]
\label{eq4cos}
\end{eqnarray}
with same wavevector length $|{\bf q}_1|=|{\bf q}_2|=|{\bf q}_3|=|{\bf q}_4|$. We have considered only the most symmetrical situation, where these wavevectors point toward the corners of a cube, corresponding to the directions $\pm(1,-1,-1),\pm(-1,1,-1),\pm(-1,-1,1)$ and $\pm(1,1,1)$. The angle between any of these wavevectors is  $70.5^\circ$. We have restricted our calculations to the case $N_{max}=3$. Our results are given in Fig.\ref{figtcfincub}, as the "cube" configuration. It is clear that this configuration is never favored. Hence when the temperature goes to zero the optimal configuration has three cosines, with wavevectors pointing in orthogonal directions. We discuss below a possible phenomenological interpretation of this result.

\subsection{Approximate $N_{max}=1$ solutions}
Since it is not so easy to find an insight into the complex solutions obtained numerically, it is worthwhile to consider the simplest meaningful approximation obtained by retaining only the lowest order Fourier components of the Green's function, that is taking $N_{max}=1$. A major advantage is that it can be handled easily analytically, which could be useful to reach a deeper understanding of the solutions for the FFLO phases. For example the $n=0$ component of the diagonal propagator is given by:
\begin{eqnarray}
g_0 = \left( 1 + 2 \sum_{i=1}^{n} |\Delta_i|^2
\frac{ \omega^2 - ({\bf k} \cdot
{\bf q}_i)^2}{\left[\omega^2 + ({\bf k} \cdot
{\bf q}_i)^2\right]^2} \right)^{-1/2}
\label{eqg0nmax1}
\end{eqnarray}
for a general order parameter, which is the sum of $n$ cosines with weights $\Delta_i$ and wavevectors ${\bf q}_i$. Naturally this makes also the final numerical treatment much easier.

\begin{figure}[htbp]
\begin{center}
\vbox to 60mm{\hspace{0mm} \epsfysize=60mm \epsfbox{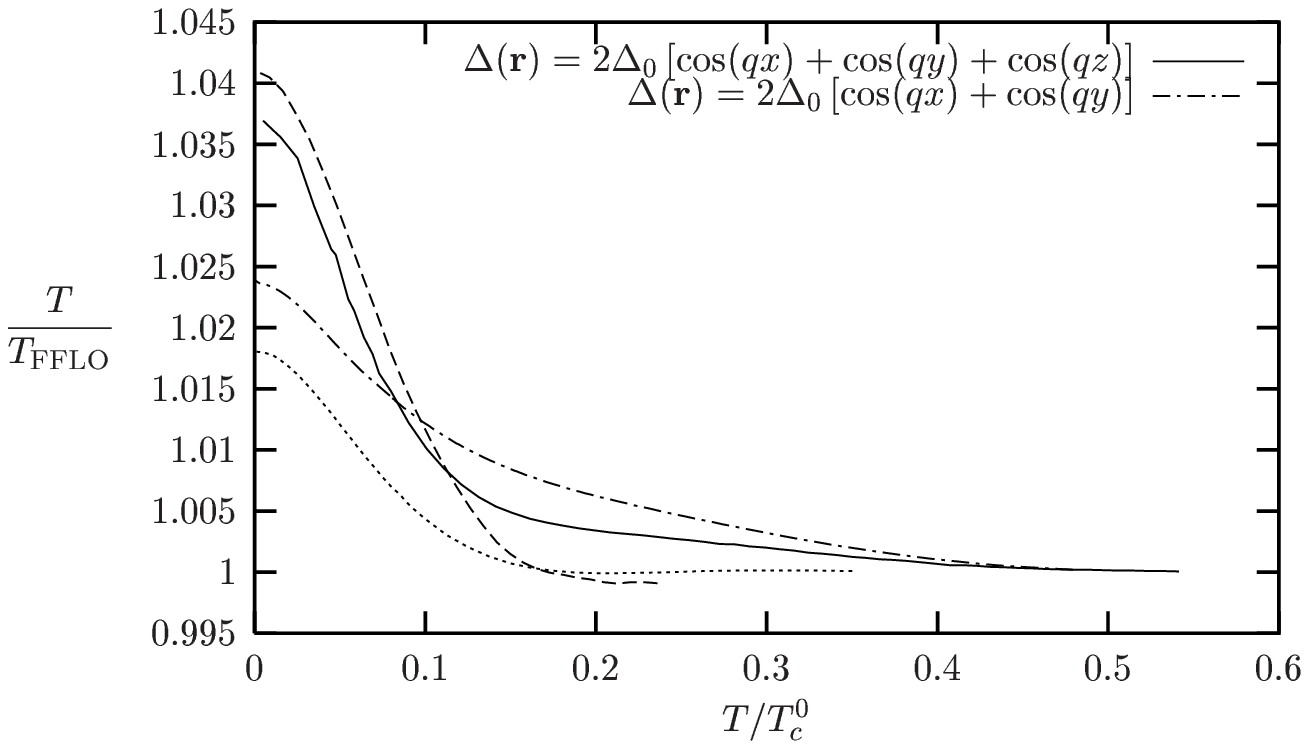}}
\vbox to 60mm{\hspace{0mm} \epsfysize=60mm \epsfbox{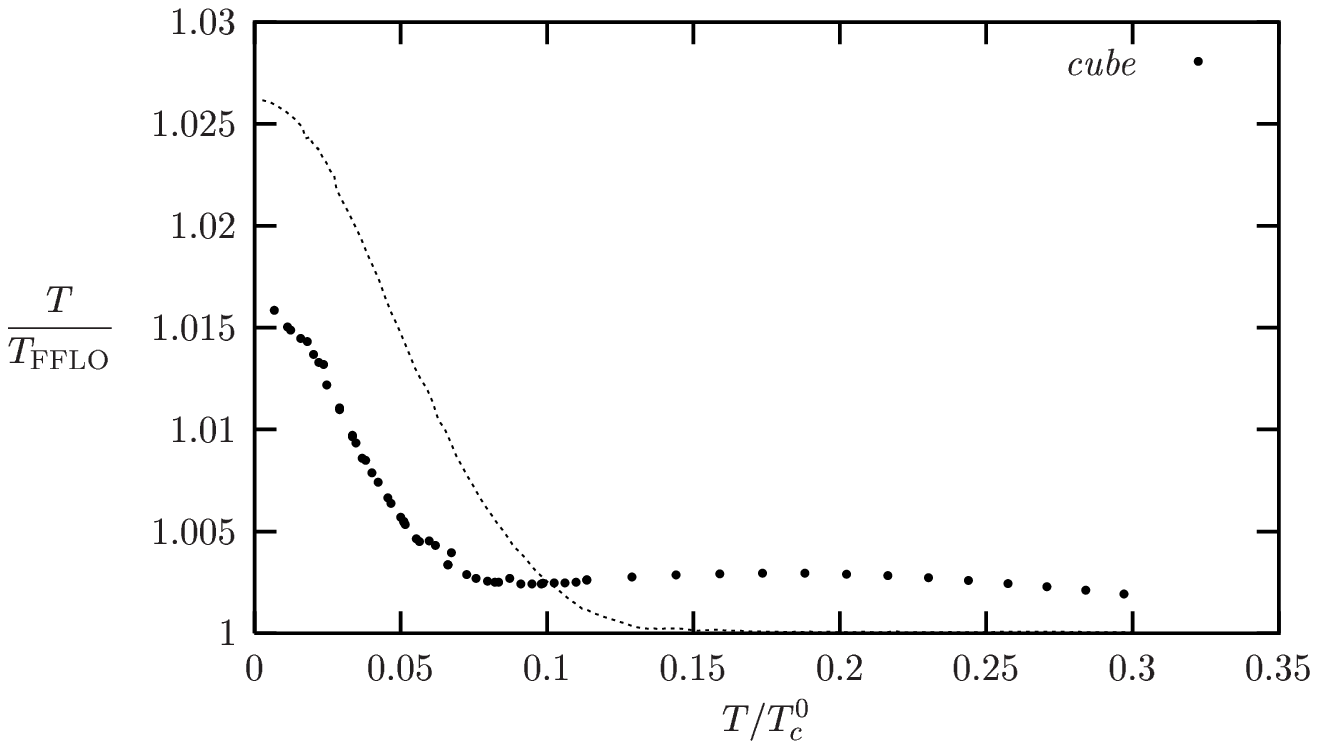}}
\caption{Upper panel: critical temperature for our two cosines order parameter with $N_{max}=1$ approximation (dotted line) compared to our complete result (dashed-dotted line). Same comparison for our three cosines order parameter : dashed line ($N_{max}=1$ approximation), full line (complete result). Lower panel : same comparison for our four cosines order parameter : dotted line ($N_{max}=1$ approximation), filled circles (complete result).}
\label{figcomp234}
\end{center}
\end{figure}

\begin{figure}[htbp]
\begin{center}
\vbox to 60mm{\hspace{0mm} \epsfysize=60mm \epsfbox{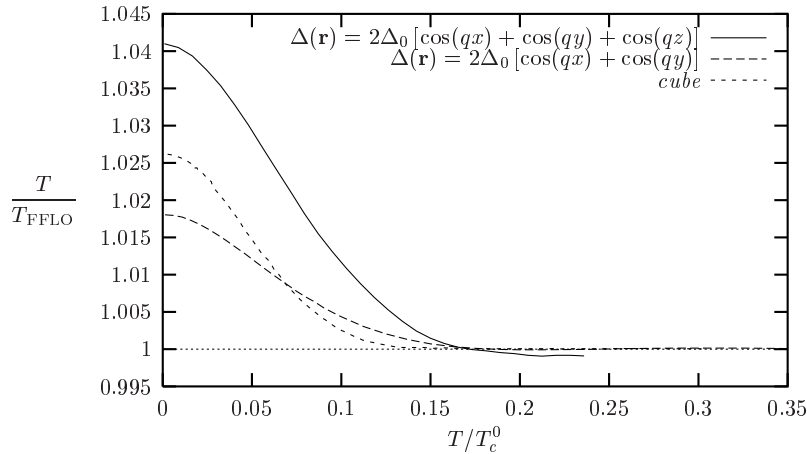}}
\caption{Comparison between the two, three and four cosines order parameters within the $N_{max}=1$ approximation.}
\label{figcompnmax1}
\end{center}
\end{figure}

In the case of a single cosine we have seen \cite{qcplana} that this solution gives the proper qualitative behaviour, with a switch from a first order to a second order transition, although the location of the switching temperature is not very accurately given. We perform here the same kind of study in the case of two, three and four cosines. In Fig.\ref{figcomp234} we compare the results of this $N_{max}=1$ solution with our complete solution, presented in the preceding subsections.
It can be seen that this $N_{max}=1$ approximation is quite satisfactory since it reproduces quite well qualitatively and semiquantitatively our complete results. Nevertheless this is compounded with the fact that all the results are quite near unity. Hence small differences, which could by themselves be considered as unsignificant, can lead to qualitative differences in the final results. This is displayed in Fig. \ref{figcompnmax1} where we gather the results for the various order parameters within the $N_{max}=1$ approximation. It can be seen that the three cosines order parameter always dominates at low temperature, in contrast to our complete results where there is a temperature range where the two cosines order parameter is the best one. Similarly the four cosines order parameter is better than the two cosines one at very low temperature within the $N_{max}=1$ approximation while this never occurs with our complete solution. In conclusion this $N_{max}=1$ approximation is quite convenient for a fast exploration of the free energy minimization problem. It gives a semiquantitatively correct picture of the competition between various order parameters. However it can not be fully trusted quantitatively.

\subsection{Phenomenological interpretation}
We compare now our results to the phenomenological picture suggested by Bowers and Rajagopal (BR) \cite{bowers}. In their $T=0$ study BR made use of a Ginzburg-Landau expansion to find the most favored order parameters in the case of a second order transition, even if in many cases the transition is actually first order which invalidates the expansion. They came in the course of this work to the following heuristic picture. To each wavevector present in the order parameter, one associates a circle drawn on the unit sphere, the axis of this circle being the wavevector direction and its angular opening $\psi_{0}$ being given by $\cos (\psi_{0}/2)=1/{\bar q}$ where ${\bar q} \simeq 1.2$ is the reduced wavevector \cite{ff,larkov} for the $T=0$ second order FFLO phase transition. This gives $\psi_{0} \simeq 67.1^{\circ}$. This circle, in the Fulde-Ferrell description, corresponds to a favored region for pairing (it is infinitely thin because we are just at the second order phase transition where the order parameter is zero). A first principle in BR heuristic picture is that the most favored order parameter has the maximum number of wavevectors, in order to increase the number of circles, i.e. the domains where favorable pairing occurs. On the other the second principle is that the crossing of two circles is very unfavorable energetically and must be avoided. This no-crossing principle of circles with a definite angular size leads clearly to an upper bound in the number of possible wavevectors. It has been found by BR to be 9 wavevectors. However the corresponding situation is not a symmetrical one for the wavevectors, and accordingly BR suggested that the symmetrical form, corresponding actually to our 'cube' configuration, would be the most favorable. This can be understood if the no-crossing principle is considered as a manifestation of an effective repulsive interaction between wavevectors directions. In this case it is reasonable to believe that the repulsion will be minimized for a symmetrical geometry for the wavevectors. 

Now we have just found that the most favorable order parameter at $T=0$ has 6 wavevectors (corresponding to our 3 cosines order parameter) instead of 8. But this can be understood qualitatively as resulting from the fact that the transition is actually first order, instead of second order, which makes the order parameter non zero at the transition. In such a case, instead of being infinitely thin, the region corresponding to favorable pairing will widen with a thickness linked to the non zero value of the order parameter. This will make the circles effectively larger, which may explain why it is no longer possible to fit 8 circles on the unit sphere and it is necessary to reduce their number to 6. Hence there is a reasonable agreement between this phenomenology and our results. We note finally that the results we have found in two dimensions \cite{mceuro} support also this picture. Here the circles on a unit sphere are replaced by to points on a unit circle. In this case the transition is second order, and as $T \rightarrow 0$ the distance between the two points goes to zero. This allows to fit on the unit circle an ever increasing number of wavevectors, as $T \rightarrow 0$. This is indeed what we have found \cite{mceuro} .

\section{The case of ultracold Fermi gases}

We will now apply our results to the case of ultracold gases, which is presently a field
of very strong interest both experimentally and theoretically. So far our analysis has been restricted to the case where the chemical potential difference between the two spin species is fixed. 
This is of course the relevant situation for 
superconductors where this chemical potential difference is produced by 
very high magnetic fields, and the resulting phase diagram which we have
found above applies directly to this case. On the other hand in ultracold
gases these are rather the populations of the various hyperfine states of the
atom under study which are under direct experimental control. Accordingly, in order to 
apply our results to the case of ultracold atomic Fermi gases, we need 
to investigate more specifically the effect of a fixed density 
difference on the phase diagram, rather than a chemical potential difference.

This modification is quite relevant in our $3$D case since we have essentially to deal
with first order transitions. This is in contrast with the $2D$ case where the transitions are
found \cite{mceuro} to be second order. Across a second order transition, 
physical quantities like the order parameter or the density are continuous.
This implies that phase diagrams for fixed chemical potentials
or fixed densities are essentially equivalent.
On the other hand the order parameter and the density are discontinuous 
for a first order transition, as for example in the well known case of the standard liquid-gas
transition. There is a forbidden domain in the phase diagram, corresponding to
phase separation, where the two phases, normal and superfluid, are coexisting in the system. This
amounts, so to speak, to split the transition line which appears for fixed chemical
potential difference into two lines in the case of fixed density difference. Each of
this line gives the density of one of the coexisting phases. In our case
the situation is somewhat more complicated since depending on the temperature
we have different phases coming into play.
Hence we consider now the density difference in the various
FFLO phases we have found.

From the solutions of the Eilenberger equations that we have
obtained using our Fourier expansion method, it is quite straightforward
to calculate any physical quantities.
In particular the difference in the densities for spin up and down is given by 
\begin{equation}\label{dendiff}
\frac{n_{\uparrow}( {\bf r}) - n_{\downarrow}( {\bf r})}{N_0} =  2 \bar{\mu} + 
4 \pi T \, {\rm Im} \sum_{n=0}^{\infty} 
< g (\omega_n -i \bar{\mu},\hat{\bf k}, {\bf r}) -  g_n (\omega_n -i \bar{\mu},\hat{\bf k})>_{\hat{\bf k}}
\end{equation}
where the first term in the right-hand side, $2 \bar{\mu}$, is actually the result for the normal state
and $g_n$ is the normal state Green's function.
It is worth noting that this density difference shows oscillations
at spatial frequencies equal to $2 \bar{q}$, $4 \bar{q}$, and so 
on, which we have found numerically to be
quite sizeable. These oscillations could be used as signatures of the FFLO phases,
in analogy with the oscillations of the magnetization in the case of superconductors.
However we will restrict ourselves to macroscopic physical quantities, and accordingly
we consider only the average value of the density difference which amounts to replace
$g$ in Eq. (\ref{dendiff}) by $g_0$.

\begin{figure}[t!]
\begin{center}
\vbox to 60mm{\hspace{0mm} \epsfysize=60mm \epsfbox{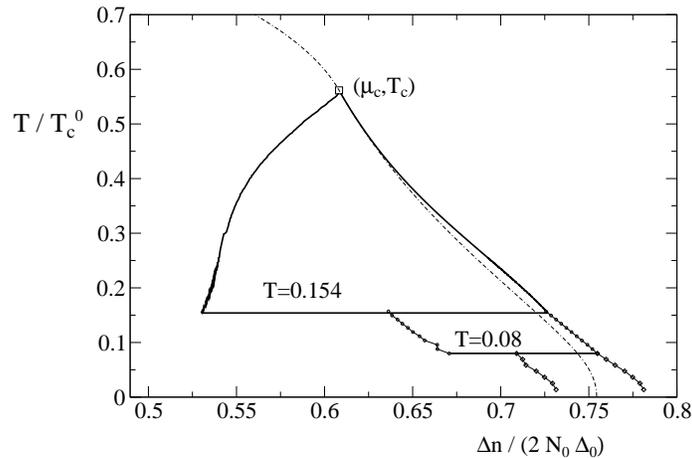}}
\caption{
Phase diagram for reduced temperature $T / T_{c0}$ versus atomic population difference $\Delta n$.
The transition lines for cases 1, 2 and 3 (see text) are respectively
represented by full lines, by full circles and by diamonds. The dashed-dotted line 
represents the standard second order FFLO transition which goes into 
the standard BCS transition above the tricritical point.
\label{phasedia}
}
\end{center}
\end{figure}

The calculation of this average density difference for the
different relevant phases leads us to the phase diagram shown
in Fig. \ref{phasedia}.

Just as in the previous case where the chemical potential was fixed,
we have three different temperature domains below the tricritical point:
\begin{enumerate}
\item The region $0.154 \, T_{c0} <T < T_{tcp} \simeq 0.561  \, T_{c0}$ where the transition is from the normal state to the superfluid state with $\Delta ({\bf r})= \Delta_1 \cos ( \bar{q} x)$.
\item The region $0.08  \, T_{c0}< T < 0.154  \, T_{c0}$ where one goes from the normal state to the 
$\Delta ({\bf r})= \Delta_1 ( \cos ( \bar{q} x) +  \cos ( \bar{q} y)) $ state. 
\item The region $T<0.08   \,T_{c0}$ where the order parameter of the superfluid state is
$\Delta ({\bf r})= \Delta_1 ( \cos ( \bar{q} x) +  \cos ( \bar{q} y) +  \cos ( \bar{q} x) )$.
\end{enumerate}
The most remarkable feature of these results is that the differences for the value of $\Delta n$ between the normal and the various FFLO phases are quite sizeable, which should allow experimentally 
to locate clearly the BCS transition. The other important point is that, between the various FFLO
phases themselves, $\Delta n$ changes in a quite noticeable way. Its evolution is qualitatively easy
to understand. The important differences in atomic populations come \cite{br} from the regions where
the order parameter is small while, in the regions where it is large, pairing favors more equal
populations. Since the number of cosines is increased when we go to lower temperatures, this
implies that the regions with small order parameter are more important. This is coherent with
the fact that, at low temperature, the superfluid phase allows a larger difference between atomic
populations than the superfluid phases found at higher temperature. This feature should allow
to locate rather easily the change of structure of the order parameter, while as we have seen the
changes in the critical temperature between the various phases is quite small and would not allow an
convenient observation.

\section{ CONCLUSION }
In this paper we have considered the original problem raised by Fulde, Ferrell, Larkin and Ovchinnikov, namely the possibility and the nature of the transition from a normal Fermi liquid to a 
superfluid phase with a space dependent order parameter, when the chemical potential of the two spin populations are different. The Fermi surface is assumed to be spherical, in three spatial dimensions. As the transition turns out to be first order we have made use of the quasiclassical Eilenberger's equations to handle this non linear problem. In order to solve these equations we have introduced a Fourier expansion for the order parameter as well as for the Green's functions. In a preceding paper \cite{qcplana} we have introduced in details this procedure, we have checked that it worked properly on the case of a one-dimensional order parameter and that it is quite efficient, at least in the vicinity of the transition. 

In the present paper we have made use of it to show that, at low temperature the order parameter switches from the one-dimensional form, found at higher temperature, which at the transition, is essentially a simple cosine, to a more complex structure. At zero temperature we find that the most stable order parameter is essentially, at the transition, the sum of three cosines with equal weights and orthogonal wavevectors. When the temperature is raised, the order parameter switches to a sum of two cosines, also with equal weights and orthogonal wavevectors, before reducing to a single cosine at higher temperature. Hence we have obtained that, at the transition, the higher Fourier components in the order parameter are essentially negligible. This makes clearly our procedure particularly efficient. We have found that, actually, the critical temperatures for the first order transitions toward these phases is only slightly higher than the critical temperature for the standard second order FFLO transition. This seems to show that there are many phases with nearly equal free energy. This should be for example included in the theory when, for superconductors, the coupling of the magnetic field to orbital degrees of freedom is also taken into account. We stress again that our method provides a way to obtain some insight in the analytical structure of this complex theory. Finally we have applied our results to the specific case of ultracold fermionic atoms and shown that the various FFLO phases arising at low temperature give rise to differences in atomic populations which allows to identify them easily.

\vspace{4mm} 
* Laboratoire associ\'e au Centre National
de la Recherche Scientifique et aux Universit\'es Paris 6 et Paris 7.

\end{document}